\documentclass[12pt]{iopart}

\def\be{\begin{equation}}
\def\te{\end{equation}}
\def\bea{\begin{eqnarray}}
\def\nn{\nonumber\\}
\def\tea{\end{eqnarray}}

\begin{document}

\title{Fourth order full quantum correlations from a Langevin-Schwinger-Dyson equation}
\author{Esteban Calzetta}
\address{ CONICET and Departamento de Fisica, FCEN}
\address{ Universidad de Buenos Aires- Ciudad Universitaria, 1428 Buenos
Aires, Argentina}
\ead{calzetta@df.uba.ar}



\begin{abstract}
It is well known that some quantum and statistical fluctuations of a quantum field may be recovered by adding suitable stochastic sources to the mean field equations derived from the Schwinger-Keldysh (Closed-time-path) effective action. In this note we show that this method can be extended to higher correlations and higher (n-particle irreducible) effective actions. As an example, we investigate three and fourth order correlations by adding stochastic sources to the Schwinger - Dyson equations derived from the 2-particle irreducible effective action. This method is a simple way to investigate the nonlinear dynamics of quantum fluctuations.
\end{abstract}
\pacs{03.70.+k, 11.10.Wx}




\section{Introduction}
Quantum fields fluctuate, and quantum fields out of equilibrium show both quantum and statistical fluctuations \cite{CH08}. In many problems of interest, the fluctuations are more relevant than the mean fields themselves. Problems that come to mind are the generation of primordial fluctuations during inflation \cite{CalHu95,CalGon97,RouVer08,WuNgFo07,WNLLC07}, the fluctuations of soft fields induced by the interaction with hard quanta \cite{GreMul97,CalHu97,BOD98,BOD99,ASY99a,ASY99b}, and the fluctuations of a Bose-Einstein condensate as described by the stochastic Gross-Pitaievskii equation \cite{ProJac08,GaAnFu01,GarDav03,BrBlGa05,Sto99,Sto01,CaHuVe07}. In such a case, one can try to obtain information about the average behavior of the fluctuations by deriving equations of motion for the fluctuation-fluctuation correlations, or else one may attempt to investigate the space-time unfolding of the fluctuations by deriving suitable Langevin-like equations for them. In certain cases, the Langevin approach is also an efficient way to derive the required self-correlations \cite{CaRoVe03}.

In the simplest set up, one is dealing with a bosonic field theory. The Heisenberg field operators are designed as $\Phi_H^a$. We use a DeWitt notation where $a$ accounts for both discrete and space-time indexes; repeated indexes are summed over the discrete ones and integrated over the continuous ones. The $\Phi_H^a$ have expectation values $\left\langle \Phi_H^a\right\rangle=\phi^a$. If only the mean fields are relevant, we may obtain causal equations of motion from them from the Schwinger-Keldysh (or closed-time-path (CTP)) 1-particle irreducible (1PI) effective action (EA). Because the Schwinger-Keldysh approach involves doubling the degrees of freedom, an extra discrete index appears, and the fields become $\Phi_H^A$ and $\phi^A$. In the simplest representation, $A=\left(i,a\right)$, where $i=1,2$ shows in which branch of the CTP we are. Other representations are also possible. The physical mean field equations, notwithstanding, are obtained from the CTP equations by adding the constraints

\be
\phi^{1a}=\phi^{2a}\equiv\phi^a
\label{ctpconstraint}
\te

If fluctuations are important, one may add noise terms to the physical mean field equations of motion. The noise self correlation is also derived from the 1PI CTP EA, more precisely from terms that do not contribute to the mean field equations when the constraints (\ref{ctpconstraint}) are enforced (see next Section). The resulting theory is still good enough to derive exact symmetric expectation values for the product of two quantum fields, that is, the Hadamard propagator 

\be
G_1^{ab}=\left\langle \left\{\Phi_H^a,\Phi_H^b\right\}\right\rangle-2\phi^a\phi^b
\label{hadamard}
\te
and in this sense it is a nontrivial extension of the mean field theory. This basic framework has been extensively used in cosmology (see \cite{CH08}; some influential papers are  \cite{CalHu94,HuMat95,LomMaz97}; see also the reviews \cite{HuVer03,Ver07}; for more recent work see \cite{HuRou07,For05,ArPaVe04,HuRoVe04,PhiHu02}) and in the theory of Bose-Einstein condensates. The important point of to which extent these fluctuations may be considered real is discussed in \cite{CH08}; for present purposes, it is enough to consider the stochastic approach as a shortcut to the actual propagators.

Sometimes higher correlations are also important. For example, one may want to compute the expectation value of the energy-momentum tensor (or the fluctuations thereof) in an interacting bosonic field theory -that usually involves three and four point functions \cite{Mot86,KuoFor93,RouVer99,PhiHu03,AnMoMo05,PeRoVe08}. Density - density correlations in a Bose-Einstein condensate are a four point correlation of the fundamental Heisenberg field; these are relevant, for example, when the condensate is investigated through Bragg scattering \cite{SOKD02,SKODTD03,PitStr03}. One may need accurate higher order order correlations to enforce important Ward-Takahashi or Slavnov-Taylor identities \cite{Ber04,CarKov08}. Or simply one may want to compute higher cumulants as a way of accounting for nonlinear effects \cite{CSHY85}. In this case, one of the most powerful computational tools is to obtain self-consistent Schwinger-Dyson equations from variations of higher n-particle irreducible (nPI) EAs where all required correlations appear on equal footing as independent variables \cite{DomMar64a,DomMar64b,NorCor75,Kim05}.

There are both practical and fundamental reasons to wish to extend the stochastic approach to field fluctuations to higher correlations as well \cite{CalHu93,CalHu95a}. On the practical side, getting the full fourth order correlations from a stochastic approach to the 2PI EA may be more efficient (or at least more heuristic) than computing the whole 4PI EA.

On the fundamental side, let us mention the following issue. It is well known that the 2PI equations of motion for the propagators lead to the Kadanoff-Baym equations for the density of states and one-particle distribution function, and eventually to the Boltzmann equation (or similar) in the appropriate limit. However, it is also known that the Boltzmann equation is only a mean field approximation to a stochastic equation, where the noise terms, in the near equilibrium case, may be derived from the fluctuation-dissipation theorem. Of course, field theory complies with the Kubo-Martin-Schwinger theorem, and therefore has the fluctuation-dissipation theorem built in. So the noise terms in the stochastic Boltzmann equation must correspond to some elements already present in the 2PI EA. The fundamental question is to make those elements explicit. This issue was solved by Hu and one of us in ref. \cite{CalHu99}. Similar issues appear at every order in the Schwinger-Dyson hierarchy.

The stochastic approach to the 1PI CTP EA exploits special features of this EA (see below) and is not readily generalizable to higher EAs. Similarly, the approach of Calzetta and Hu in ref. \cite{CalHu99} is also ad hoc, in this case for the 2PI EA. A systematic framework, which could be applied to any EA and to the symmetry broken or unbroken cases alike, would be highly desirable, not least because of the light it sheds on the particular approaches devised for the 1PI and 2PI cases. Our aim in this note is to develop such an uniform formalism.

The rest of the paper is organized as follows.

In the next section we study noise in the 1PI theory. We first show how one can associate a CTP 1PI EA efective action to a problem defined in terms of a Langevin equation. We then show that the 1PI EA for a quantum field theory problem has, under certain approximations, the same structure as the EA arising from a stochastic problem. This allows the direct identification of the equivalent stochastic problem to a given field theory. As an application, we review the derivation of the Hadamard propagator of the full theory from the equivalent stochastic equation. 

In the following Section we review the 2PI EA and two early attempts of a stochastic formulation of the propagator dynamics \cite{CalHu95a,CalHu99}. We show the shortcomings of these attempts and how they differ from the proposal in this note.

Finally we present a systematic approach to building stochastic equivalents for a given field theory and apply it to the 1PI and 2PI cases. Although we shall not discuss it explicitly, generalization to higher effective actions is straightforward. In the 2PI case, we finally obtain the same result as in \cite{CalHu99}, but without the contrived arguments contained in that paper. We show the basic oversight contained in  \cite{CalHu99}, which obscured the simple derivation of the 2PI noise presented here.

The paper ends with some brief final remarks.

\section{Stochastic approach to the 1PI EA}
The goal of this Section is to provide an heuristic introduction to stochastic equations derived from the 1PI EA. For a deeper discussion see \cite{CH08,GRLE98}.

\subsection{From Langevin equations to effective actions}
To see why it is natural to translate a problem described in terms of an effective action into an equivalent Langevin equation framework, let us first traverse the opposite road, that is, how to associate a generating functional to a problem whose primary description is in terms of a stochastic equation of motion.

We therefore assume we have a string of c-number fields $\Phi_s^a$ obeying a system of equations of the form \cite{MaSiRo73,CooRos01,ZanCal02,ChaDem01a,ChaDem01b}

\be
\mathbf{D}_a\left[\Phi_s\right]=-q_{b}F^b_a\left[\Phi_s\right]
\label{ne1}
\te
where $\mathbf{D}_a$ and $F^b_a$ are possibly nonlinear functionals and the $q_{b}$ are stochastic Gaussian variables with zero mean and self-correlation $\left\langle q_{b}q_{c}\right\rangle =Q_{bc}$. To avoid the complexities associated to nonlinear Langevin equations, we take this equation to mean that it is possible to write $\Phi_s^a=\phi^a+\varphi^a$, where
$\phi^a$ is a solution of the homogeneous equation

\be
\mathbf{D}_a\left[\phi\right]=0
\label{ne2}
\te
and

\be
\mathbf{D}_{a,c}\left[\phi\right]\varphi^c=-q_{b}F^b_a\left[\phi\right]
\label{ne3}
\te
where here and henceforth a comma denotes a functional derivative

\be
\mathbf{D}_{a,c}\equiv\frac{\delta \mathbf{D}_a}{\delta\phi^c}
\label{ne4}
\te
If there are no zero modes, this implies that the expectation value $\left\langle \varphi^a\right\rangle=0$, so within this approximation we may say that $\phi^a$ is the expectation value of $\Phi_s^a$.

The generating functional for expectation values of the $\Phi_s^a$ fields is

\be
Z\left[J\right]\equiv e^{iW\left[J\right]}=\int\:D\Phi_s\:Dq\:\mathcal{Q}\:\delta\left(\Phi_s^a-\phi^a-G_{ret}^{ab}q_{c}F^c_b\left[\phi\right]\right)e^{iJ_d\Phi_s^d}
\label{ne5}
\te
where $G_{ret}^{ab}$ is the causal Green function for the operator $\mathbf{D}_{a,c}$, namely

\be
\mathbf{D}_{a,c}G_{ret}^{cb}=-\delta^b_a
\label{ne6}
\te
and $\mathcal{Q}$ is the Gaussian probability density functional of the $q$ sources. 
The generating functional obeys

\be
W\left[0\right]=0
\label{ne6b}
\te

\be
W^{,a}\left[0\right]=\left.\frac{\delta W}{\delta J_a}\right|_{J=0}=\phi^a
\label{ne6c}
\te

Shifting the $\Phi_s^a$ fields by an amount $\phi^a$, namely  $\Phi_s^a=\phi^a+\varphi_+^a$, we get

\be e^{iW\left[J\right]}=e^{iJ_d\phi^d}\int\:D\varphi_+\:Dq\:\mathcal{Q}\:\delta\left(\varphi_+^a-G_{ret}^{ab}q_{c}F^c_b\left[\phi\right]\right)e^{iJ_d\varphi_+^d}
\label{ne7}
\te
This may be rewritten as

\be 
e^{iW\left[J\right]}=\frac{e^{iJ_d\phi^d}}{Det G_{ret}}\int\:D\varphi_+\:Dq\:\mathcal{Q}\:\delta\left(\mathbf{D}_{a,c}\varphi_+^c+q_{c}F^c_a\left[\phi\right]\right)e^{iJ_d\varphi_+^d}
\label{ne8}
\te
We now introduce an auxiliary field $\varphi_-$ to exponentiate the delta function

\be 
e^{iW\left[J\right]}=\frac{e^{iJ_d\phi^d}}{Det G_{ret}}\int\:D\varphi_+\:D\varphi_-\:Dq\:\mathcal{Q}\:e^{i\varphi_-^a\left(\mathbf{D}_{a,c}\varphi_+^c+q_{c}F^c_a\left[\phi\right]\right)}e^{iJ_d\varphi_+^d}
\label{ne9}
\te
and perform the functional integral over the sources

\be 
e^{iW\left[J\right]}=\frac{e^{iJ_d\phi^d}}{Det G_{ret}}\int\:D\varphi_+\:D\varphi_-\:e^{i\varphi_-^a\mathbf{D}_{a,c}\varphi_+^c-\frac12\varphi_-^aN_{ab}\varphi_-^b}e^{iJ_d\varphi_+^d}
\label{ne10}
\te
where

\be
N_{ab}=Q_{cd}F^c_a\left[\phi\right]F^d_b\left[\phi\right]
\label{ne11}
\te
This generating functional is the particular case, when $J^+=J$ and $J^-=0$, of the functional

\be 
e^{iW\left[J^+,J^-\right]}=\frac{e^{iJ^+_d\phi^d}}{Det G_{ret}}\int\:D\varphi_+\:D\varphi_-\:e^{i\varphi_-^a\mathbf{D}_{a,c}\varphi_+^c-\frac12\varphi_-^aN_{ab}\varphi_-^b}e^{i\left[J^+_d\varphi_+^d+J^-_d\varphi_-^d\right]}
\label{ne10b}
\te
Let us write

\be
W\left[J^+,J^-\right]=J^+_d\phi^d+w\left[J^+,J^-\right]
\label{ne11b}
\te
$w\left[J^+,J^-\right]$ is obtained from a functional Fourier transform of a Gaussian functional, and as such it is a quadratic function. It therefore obeys the Euler theorem

\be
w\left[J^+,J^-\right]=\frac12\left\{J^+_d\bar\varphi_+^d+J^-_d\phi_-^d\right\}
\label{ne12}
\te
where we have introduced two new background fields

\be
\bar\varphi_+^d=\frac{\delta w}{\delta J^+_d}
\label{ne12b}
\te

\be
\bar\phi_-^d=\frac{\delta w}{\delta J^-_d}
\label{ne14}
\te
The expectation value of the $+$ field in the presence of non zero external sources is
\be
\phi_+^d=\frac{\delta W}{\delta J^+_d}=\phi^d+\bar\varphi_+^d
\label{ne15}
\te

We can now compute the effective action as the Legendre transform of the generating functional

\be
\Gamma\left[\phi_+,\phi_-\right]=W\left[J^+,J^-\right]-\left\{J^+_d\phi_+^d+J^-_d\phi_-^d\right\}=\frac{-1}2\left\{J^+_d\bar\varphi_+^d+J^-_d\phi_-^d\right\}
\label{ne16}
\te
We have the identities

\be
\phi_-^a\mathbf{D}_{a,d}=-J^+_d
\label{ne17}
\te

\be
\mathbf{D}_{d,c}\bar\varphi_+^c+iN_{db}\phi_-^b=-J^-_d
\label{ne18}
\te
and therefore

\be
\Gamma\left[\phi_+,\phi_-\right]=\phi_-^a\mathbf{D}_{a,d}\bar\varphi_+^d+\frac{i}2\phi_-^dN_{db}\phi_-^b
\label{ne19}
\te
or else, adding $\mathbf{D}_a\left[\phi\right]=0$

\be
\Gamma\left[\phi_+,\phi_-\right]=\phi_-^a\mathbf{D}_a\left[\phi_+\right]+\frac{i}2\phi_-^dN_{db}\left[\phi_+\right]\phi_-^b
\label{ne20}
\te
This shows how we can associate an effective action to a Langevin type equation.

We finally mention the self correlation for the $\varphi^a$ fields. Since we have the explicit representation $\varphi^a=G_{ret}^{ab}q_{c}F^c_b$ we get

\be
\left\langle \varphi^a\varphi^b\right\rangle \equiv G_s^{ab}=G_{ret}^{ac}F^d_cG_{ret}^{be}F^f_eQ_{df}=G_{ret}^{ac}G_{ret}^{be}N_{ce}
\label{ne21}
\te

\subsection{From effective actions to Langevin equations }
Let us now investigate a scalar field theory described by the Heisenberg operators $\Phi_H^a$. The fields have one and two-particle expectation values

\be
\left\langle \Phi_H^a\right\rangle=\phi^a
\te

\be
\left\langle \Phi_H^a\Phi_H^b\right\rangle=\phi^a\phi^b+G^{ab}
\te
These expectation values cannot be derived either from the euclidean generating functional or its analytic continuation to Minkowski space, which generate IN-OUT matrix elements instead \cite{CH08}. To find  a suitable generating functional, we must consider two external sources $J^A=\left(J^{1a},J^{2a}\right)$ and introduce the Schwinger - Keldysh or closed time-path (CTP) 1 particle-irreducible (1PI) generating functional
$W_{1PI}\left[J_A\right]$ as

\be
Z_{1PI}\left[J\right]=e^{iW_{1PI}\left[J\right]}=\left\langle \left(\tilde{T}\left[e^{-iJ^{2}_a\Phi_H^a}\right]\right)\left(T\left[e^{iJ^{1}_b\Phi_H^b}\right]\right)\right\rangle
\label{ne22}
\te
Where $T$ ($\tilde{T}$) means (anti) temporal ordering. We may now derive the expectation value

\be
\phi^a=\left.\frac{\delta W_{1PI}}{\delta J^1_a}\right|_{J^1=J^2=0}
\label{ne23}
\te
More generally, we may consider the expectation value of the Heisenberg operator driven by an external source $J_a$. This is

\be
\phi^a=\left.\frac{\delta W_{1PI}}{\delta J^1_a}\right|_{J^1=J^2=J}
\label{ne24}
\te
Even more generally, we may consider this as the theory of a field doublet $\Phi^A=\left(\Phi^{1a},\Phi^{2a}\right)$ coupled to sources $J_A=\left(J_{1a}=J^{1a},J_{2a}=-J^{2a}\right)$ (observe that lowering or raising a $2$ index involves a sign change). In this theory we have two backgrounds fields

\be
\phi^A=\frac{\delta W_{1PI}}{\delta J_A}\equiv W_{1PI}^{,A}
\label{ne25}
\te
but the physical situation is when

\be
\phi_-^a=\phi^{1a}-\phi^{2a}=0
\label{ne26}
\te
When this obtains, then

\be
\phi_+^a=\frac12\left(\phi^{1a}+\phi^{2a}\right)=\phi^a
\label{ne27}
\te
is the physical expectation value

The 1PI effective action is the Legendre transform

\be
\Gamma_{1PI}\left[\phi^A\right]=W_{1PI}-J_A\phi^A
\te
Therefore

\be
\Gamma_{1PI,A}=\frac{\delta \Gamma_{1PI}}{\delta \phi^A}=-J_A
\label{mean}
\te

The 1PI CTP EA can be written generically as \cite{CH08}

\begin{equation}
\Gamma \left[ \phi_{-},\phi_{+}\right] =\phi_{-}^a
\mathbf{D}_a\left[ \phi_{+}\right] +\frac{i}{2}\phi_{-}^a \mathbf{N}_{ab}\left[ \phi_{+}\right]
\phi_{-}^b +...  \label{d80}
\end{equation}
Where the ellipsis means terms of higher order in $\phi_-$. Here $\mathbf{D}$ ($\mathbf{N}$) is the so-called dissipation (noise) kernel. This is of course identical in form to the effective action for a stochastic theory \ref{ne20}, and thereby we may write down the equivalent Langevin equation

\be
\mathbf{D}_a\left[ \Phi_s\right] =-\tilde{\xi}_a 
\label{classpath}
\te
where the $\tilde{\xi}_a $ are Gaussian stochastic sources with zero mean and self-correlation

\be
\left\langle\tilde{\xi}_{a}\tilde{\xi}_{b} \right\rangle=\mathbf{N}_{ab}
\te

\subsection{Full Hadamard propagator from the stochastic 1PI approach}
As an application of the stochastic 1PI approach we shall show that the two point function for the stochastic theory \ref{ne21} is identical to half the Hadamard propagator of the full quantum theory.

The full quantum propagators may be derived from the 1PI generating functional as
\be
G^{AB}=-i\frac{\delta^2 W_{1PI}}{\delta J_A\delta J_B}\equiv -iW_{1PI}^{,AB}
\te
From the properties of the Legendre transform we have

\be
\Gamma_{1PI,AB}W_{1PI,}^{BC}=-\delta^C_A
\te
whereby

\be
\Gamma_{1PI,AB}G^{BC}=i\delta^C_A
\te
In the $A=\left(\alpha ,a\right)$ representation, where $\alpha =\pm$, we find

\be
\Gamma_{1PI,\left(\alpha ,a\right)}=
\left(\begin{array}{c}
	\phi_{-}^c\left[\mathbf{D}_{c,a}+\frac{i}{2}\mathbf{N}_{cb,a}
\phi_{-}^b\right]\\
\mathbf{D}_a\left[ \phi_{+}\right]+i\mathbf{N}_{ab}
\phi_{-}^b
\end{array}\right)
\te
And therefore the Hessian, evaluated at the physical point $\phi_{-}^c=0$ is

\be
\Gamma _{1PI,\left(\alpha ,a\right),\left(\beta ,b\right)}=
\left(\begin{array}{cc}
	0&\mathbf{D}_{b,a}\\
\mathbf{D}_{a,b}&i\mathbf{N}_{ab}
\end{array}\right)
\label{Hessian}
\te
We also identify

\be
G^{\left(\alpha ,a\right),\left(\beta ,b\right)}=
\left(\begin{array}{cc}
\frac12G_1^{ab}&-iG_{ret}^{ab}\\
-iG_{adv}^{ab}&0
\end{array}\right)
\label{onshellprop}
\te
where $G_{ret}^{ab}$ is the retarded propagator, $G_{adv}^{ab}=G_{ret}^{ba}$ is the advanced propagator, and $G_1^{ab}$ is the Hadamard propagator

\be
G_1^{ab}=\left\langle \left\{\Phi_H^a,\Phi_H^b\right\}\right\rangle
\label{ne50}
\te
The  equations for the propagators become

\be
\left(\begin{array}{cc}
	0&\mathbf{D}_{b,a}\\
\mathbf{D}_{a,b}&i\mathbf{N}_{ab}
\end{array}\right)
\left(\begin{array}{cc}
\frac12G_1^{bc}&-iG_{ret}^{bc}\\
-iG_{adv}^{bc}&0
\end{array}\right)
=i
\left(\begin{array}{cc}
\delta^c_a&0\\
0&\delta^c_a
\end{array}\right)
\label{onshelleq}
\te
namely

\bea
\mathbf{D}_{a,b}G_{ret}^{bc}&=&-\delta^c_a\nn
\mathbf{D}_{b,a}G_{adv}^{bc}&=&-\delta^c_a\nn
\mathbf{D}_{a,b}G_1^{bc}&=&-2\mathbf{N}_{ab}G_{adv}^{bc}
\tea
This last equation shows that indeed $G_1$ is twice the stochastic propagator $G_s^{ab}$ defined in eq. \ref{ne21}.

\subsection{Remarks}
In this Section we have shown that the 1PI EA derived from a quantum field theory, cut off at terms quadratic in the difference field $\phi_-$, is identical in form to the effective action derived from a suitable Langevin equation. This allows us, for example, to compute the full Hadamard function from the stochastic approach.

It would be desirable to extend this equivalence to higher correlations, and to this effect we may wish to find stochastic equivalents to the higher effective actions to be introduced in next Section. However, the stochastic approach as presented above relies heavily on the representation \ref{d80} for the 1PI EA. The structure of the 1PI EA is not replicated in the higher effective actions (except under restrictive conditions on the propagators, see \cite{CalHu95a}) and so this method fails in general.

In the following we shall present an alternative derivation of the stochastic approach which does not rely in any particular feature of the 1PI EA, and therefore it is readily generalized to higher effective actions. As a first step, we shall briefly introduce the 2PI effective action.

\section{The 2PI EA and early stochastic formulations}

\subsection{The 2PI EA}
The 1PI generating functional introduced in last Section admits a path integral representation in terms of fields defined on the closed time-path

\be
e^{iW_{1PI}\left[J_A\right]}=\int\:D\Phi^A\:e^{i\left[S\left[\Phi^A\right]+J_A\Phi^A\right]}
\label{ne60}
\te
where $S\left[\Phi^A\right]=S\left[\Phi^1\right]-S\left[\Phi^2\right]^*$ is the classical closed time-path action. 
In the 2PI theory we add  nonlocal sources $K_{AB}$ coupled to $\left(1/2\right)\Phi_H^A\Phi_H^B$. Thus the CTP 2PI generating functional is

\be
e^{iW\left[J_A,K_{AB}\right]}=\int\:D\Phi^A\:e^{i\left[S\left[\Phi^A\right]+J_A\Phi^A+\frac12K_{AB}\Phi^A\Phi^B\right]}
\label{ne61}
\te

The first and second derivatives of the 2PI generating potential $W$ read

\be
W^{,A}=\phi^A
\te

\be
W^{,\left(AB\right)}=\frac12\left[\phi^A\phi^B+G^{AB}\right]
\te

\be
W^{,AB}=iG^{AB}
\te

\be
W^{,A\left(BC\right)}=\frac i2\left[\left\langle \Phi_H^A\Phi_H^B\Phi_H^C\right\rangle-\phi^A\left(\phi^B\phi^C+G^{BC}\right)\right]
\te

\be
W^{,\left(AB\right)\left(CD\right)}=\frac i4\left[\left\langle \Phi_H^A\Phi_H^B\Phi_H^C\Phi_H^D\right\rangle-\left(\phi^A\phi^B+G^{AB}\right)\left(\phi^C\phi^D+G^{CD}\right)\right]
\te
Observe that in these equations we are writing

\be
W^{,\left(AB\right)}=\frac{\delta W}{\delta K_{AB}}
\te
to distinguish it from

\be
W^{,AB}=\frac{\delta^2 W}{\delta J_A\delta J_B}
\te
The 2PI CTP EA is the full Legendre transform

\be
\Gamma =W-J_A\phi^A-\frac12K_{AB}\left[\phi^A\phi^B+G^{AB}\right]
\te
Therefore the mean field equations of motion are

\be
\Gamma_{,A}=-J_A-K_{AB}\phi^B
\te

\be
\Gamma_{,\left(AB\right)}\equiv\frac{\delta \Gamma}{\delta G^{AB}}=-\frac12K_{AB}
\te

One further variation yields the identities

\be
\left[\Gamma_{,AC}-2\Gamma_{,\left(AC\right)}\right]\phi^{C,E}+\Gamma_{,A\left(CD\right)}G^{CD,E}=-\delta_{A}^E
\te

\be
\left[\Gamma_{,AC}-2\Gamma_{,\left(AC\right)}\right]\phi^{C,\left(EF\right)}+\Gamma_{,A\left(CD\right)}G^{CD,\left(EF\right)}=-\delta_{\left(AB\right)}^{\left(EF\right)}\phi^B
\te

\be
\Gamma_{,\left(AB\right)C}\phi^{C,E}+\Gamma_{,\left(AB\right)\left(CD\right)}G^{CD,E}=0
\te

\be
\Gamma_{,\left(AB\right)C}\phi^{C,\left(EF\right)}+\Gamma_{,\left(AB\right)\left(CD\right)}G^{CD,\left(EF\right)}=-\frac12\delta_{\left(AB\right)}^{\left(EF\right)}
\te
where $\delta_{\left(AB\right)}^{\left(EF\right)}$ stands for the symmetrized identity operator

\be
\delta_{\left(AB\right)}^{\left(EF\right)}=\frac12\left\{\delta_A^E\delta_B^F+\delta_A^F\delta_B^E\right\}
\te

We now write the derivatives of the mean fields in terms of correlations

\be
\phi^{C,E}=W^{,CE}=iG^{CE}
\te

\be
\phi^{C,\left(EF\right)}=W^{,C\left(EF\right)}=\frac i2\left[\left\langle \Phi_H^C\Phi_H^E\Phi_H^F\right\rangle-\phi^C\left(\phi^E\phi^F+G^{EF}\right)\right]
\te

\bea
G^{CD,E}&=&2W^{,\left(CD\right)E}-\phi^{C,E}\phi^D-\phi^{D,E}\phi^C\nn
&=&i\left[\left\langle \Phi_H^C\Phi_H^D\Phi_H^E\right\rangle-\phi^E\phi^C\phi^D-\phi^EG^{CD}-\phi^CG^{DE}-\phi^DG^{CE}\right]
\tea

\bea
G^{CD,\left(EF\right)}&=&2W^{,\left(CD\right)\left(EF\right)}-\phi^{C,\left(EF\right)}\phi^D-\phi^{D,\left(EF\right)}\phi^C\nn
&=&\frac i2\left\{\left\langle \Phi_H^C\Phi_H^D\Phi_H^E\Phi_H^F\right\rangle-\left(\phi^C\phi^D+G^{CD}\right)\left(\phi^E\phi^F+G^{EF}\right)\right.\nn
&-&\phi^D\left[\left\langle \Phi_H^C\Phi_H^E\Phi_H^F\right\rangle-\phi^C\left(\phi^E\phi^F+G^{EF}\right)\right]\nn
&-&\left.\phi^C\left[\left\langle \Phi_H^D\Phi^E_H\Phi^F_H\right\rangle-\phi^D\left(\phi^E\phi^F+G^{EF}\right)\right]\right\}
\tea
This last equation may be rewritten as

\bea
G^{CD,\left(EF\right)}&=&\frac i2\left\{\left\langle \Phi_H^C\Phi_H^D\Phi_H^E\Phi_H^F\right\rangle-\phi^D\left\langle \Phi_H^C\Phi_H^E\Phi_H^F\right\rangle-\phi^C\left\langle \Phi_H^D\Phi_H^E\Phi_H^F\right\rangle\right.\nn&-&\left.
\left(-\phi^C\phi^D+G^{CD}\right)\left(\phi^E\phi^F+G^{EF}\right)\right\}
\tea
and further

\bea
-2iG^{CD,\left(EF\right)}&=&\left\langle \Phi_H^C\Phi_H^D\Phi_H^E\Phi_H^F\right\rangle\nn
&-&\phi^C\phi^D\phi^E\phi^F\nn
&-&\phi^C\phi^DG^{EF}-\phi^C\phi^EG^{DF}-\phi^C\phi^FG^{DE}\nn&-&\phi^D\phi^EG^{CF}-\phi^D\phi^FG^{CE}-\phi^E\phi^FG^{CD}\nn
&+&i\phi^CG^{EF,D}+i\phi^DG^{EF,C}
\tea

We also notice the identity

\be
-2i\phi^{C,\left(EF\right)}=-iG^{CE,F}+\phi^EG^{CF}+\phi^FG^{CE}
\te

\subsection{Early stochastic formulations}
As we have said in the Introduction, it has been known for a long time that the Boltzmann equation is just a mean field equation, and can be improved by upgrading it to a full Langevin type equation where particle number fluctuations are explicitly included. Since from the point of view of field theory the Boltzmann equation is just a particular limit of the Kadanoff-Baym equations, which are in turn equivalent to the Schwinger-Dyson equations \cite{CH08}, it is natural to seek a corresponding stochastic generalization of the latter. In this subsection we shall review two early attempts in this direction and point out their shortcomings. The proper stochastic formulation shall be presented in next Section.

As we have said in Section II, the peculiar structure of the 1PI EA allows to associate to a field theory problem an equivalent Langevin equation, whereby, for example, the Hadamard propagator may be obtained as a stochastic average. One possible strategy is to try to cast the 2PI EA in a similar framework. This line of thought is pursued in \cite{CalHu95a}. Success is found only under special, and restrictive, assumptions on the structure of the propagators, and therefore it is unsuitable as a general foundation for the formalism.

In \cite{CalHu99} the same authors follow a different strategy, which may be regarded as a nonequilibrium generalization of the fluctuation - dissipation theorem. We shall consider here only the case of a field theory with no background mean fields. The general case will be discussed in next Section.

We assume there are stochastic kernels $G_s^{AB}=G^{AB}+\gamma^{AB}$ such that the stochastic averages of these kernels reproduce the quantum averages of products of the composite operator $\Phi^A\Phi^B$. These kernels obey the Langevin equation

\be
\Gamma_{,(AB)}\left[G_s\right]=\frac{-1}2\kappa_{AB}
\label{ne72}
\te
After linearization, this becomes

\be
\Gamma_{,(AB)(CD)}\gamma^{CD}=\frac{-1}2\kappa_{AB}
\label{ne73}
\te

We have two ways of computing the self correlation for the stochastic kernels. By assumption

\be
\left\langle G_s^{AB}G_s^{CD}\right\rangle =\left\langle \Phi_H^A\Phi_H^B\Phi_H^C\Phi_H^D\right\rangle=G^{AB}G^{CD}-4iW^{,(AB)(CD)}
\label{ne74}
\te
while from the explicit solution of the linearized equations we get

\bea
\left\langle G_s^{AB}G_s^{CD}\right\rangle &=&G^{AB}G^{CD}+\left\langle \gamma^{AB}\gamma^{CD}\right\rangle\nn &=&G^{AB}G^{CD}+\frac14 \left[\Gamma_{,(AB)(EF)}\right]^{-1}\left[\Gamma_{,(CD)(GH)}\right]^{-1}\left\langle \kappa_{EF}\kappa_{GH}\right\rangle
\label{ne75}
\tea
Asking both computations to agree we get

\be
\left\langle \kappa_{IJ}\kappa_{KL}\right\rangle=-16i\Gamma_{,(IJ)(AB)}\Gamma_{,(KL)(CD)}W^{,(AB)(CD)}=4i\Gamma_{,(IJ)(KL)}
\label{ne76}
\te
As we shall see in next Section, this is the correct result. However, the authors of \cite{CalHu99}  provide some contrived argument to the effect that the sign of the last term in eq. (\ref{ne76}) ought to be changed. These arguments would be unconvincing, except the sign reversal seems to be necessary to derive the correct noise term for the Boltzmann - Langevin equation in the kinetic theory limit.

This seeming paradox demands clarification, which will be provided in next Section.

\section{Systematic Langevin approach}

\subsection{Systematic Langevin approach to mean field dynamics}

So far in this note we have seen two different strategies to associate an equivalent stochastic equation to a field theory problem. The approach of Section 2 depends upon the formal analogy between the 1PI EA for the quantum and stochastic problems. This approach is straightforward for the 1PI theory, but it is not easily generalizable to higher effective actions. The approach of Section 3 assumes there is an equivalent stochastic description, where the noise is whatever it needs to be to sustain the proper fluctuations. In this sense, it is a generalized fluctuation - dissipation kind of argument. This approach works for any effective action, but seems to be subject to a sign ambiguity. In this Section we shall develop this second approach in a systematic  way, dispelling the apparent paradox afflicting the discussion in  \cite{CalHu99}.

Let us reformulate the 1PI problem in terms of the fluctuation - dissipation inspired approach.
To derive the Langevin-like approach, we write down eq. (\ref{mean}) for a stochastic c-number CTP field variable $\Phi_s^A$ coupled to a \emph{stochastic} source $\xi^{1PI}_A$. We assume the solution takes the form 

\be
\Phi_s^A=\phi^A+\varphi^A
\te
where the displacements $\varphi^A$ account for both the quantum and statistical fluctuations

\be
\left\langle \varphi^A\varphi^B\right\rangle =G^{AB}
\te
In the linearized regime, we get

\be
\Gamma_{1PI,AB}\varphi^B=-\xi^{1PI}_A
\label{1pilan}
\te
We find

\be
\varphi^B=iG^{BC}\xi^{1PI}_C
\te

\be
\left\langle \varphi^A\varphi^B\right\rangle =-G^{AC}G^{BD}\left\langle \xi^{1PI}_C\xi^{1PI}_D\right\rangle 
\te
so we require, using that $G^{BD}=G^{DB}$

\be
\left\langle \xi^{1PI}_C\xi^{1PI}_D\right\rangle G^{DB}=-\delta^B_C
\te
that is

\be
\left\langle \xi^{1PI}_C\xi^{1PI}_D\right\rangle =i\Gamma_{1PI,CD}
\label{1pinoise}
\te
It is important to stress the differences between this approach and that of Section 2. In Section 2 there was only one stochastic source coupled to the stochastic condensate field. In the new approach of this section, the number of stochastic sources has been doubled. 

Other way to express the same concept is to observe that the constraint $\phi_-=0$ which characterizes physical mean field configurations is violated in our stochastic formulation. 
The formal stochastic equations for the fluctuations read

\bea
\varphi^{-b}\mathbf{D}_{b,a}&=&-\xi^{1PI}_{+a}\nn
\mathbf{D}_{a,b}\varphi^{+b}+i\mathbf{N}_{ab}\varphi^{-b}&=&-\xi^{1PI}_{-a}
\tea
The first equation yields

\be
\varphi^{-b}=G_{adv}^{bc}\xi^{1PI}_{+c}
\te
We can now eliminate $\varphi^{-b}$ and obtain a single stochastic equation for $\varphi^{+b}$. The question arises if this single equation is the same as derived in Section 2. This requires  we identify

\be
\tilde{\xi}_a=\xi^{1PI}_{-a}+i\mathbf{N}_{ab}G_{adv}^{bc}\xi^{1PI}_{+c}
\te
Now we have $\left\langle\xi^{1PI}_{-a}\xi^{1PI}_{-b} \right\rangle =-\mathbf{N}_{ab}$, $\left\langle\xi^{1PI}_{-a}\xi^{1PI}_{+b} \right\rangle =i\mathbf{D}_{a,b}$ and $\left\langle\xi^{1PI}_{+a}\xi^{1PI}_{+b} \right\rangle =0$, so the above identification implies

\be
\left\langle\tilde{\xi}_{a}\tilde{\xi}_{b} \right\rangle =-\mathbf{N}_{ab}-\mathbf{N}_{bd}G_{adv}^{dc}\mathbf{D}_{a,c}-\mathbf{N}_{ad}G_{adv}^{dc}\mathbf{D}_{b,c}=\mathbf{N}_{ab}
\te
as expected. After the auxiliary field $\varphi^{-b}$ has been integrated out, both stochastic formulations are equivalent.

\subsection{The stochastic 2PI theory}

One advantage of the new formalism is that it can be trivially generalized to higher effective actions. Let us consider again the 2PI theory.

This time we shall consider the general case where symmetry may be broken. To this end we write as before $\Phi_s^A=\phi^A+\varphi^A$, but now we also write a stochastic representation for the composite operator $\left(\Phi_H^A-\phi^A\right)\left(\Phi_H^B-\phi^B\right)$. In the stochastic theory this composite operator is represented by a stochastic c-number kernel $G_s^{AB}=G^{AB}+\gamma^{AB}$. We assume the identity

\be
\Phi_s^A\Phi_s^B=\phi^A\phi^B+G^{AB}+\phi^A\varphi^B+\varphi^A\phi^B+\gamma^{AB}
\te
The required expectation values are

\be
\left\langle \Phi_H^C\Phi_H^E\right\rangle=\phi^C\phi^E+\left\langle \varphi^C\varphi^E\right\rangle
\te

\bea
\left\langle \Phi_H^C\Phi_H^D\Phi_H^E\right\rangle&=&\left\langle \Phi_H^C\left(\Phi_H^D\Phi_H^E\right)\right\rangle\nn
&=&\phi^C\phi^D\phi^E+\phi^CG^{DE}+\phi^DG^{CE}+\phi^EG^{CD}+\left\langle \varphi^C\gamma^{DE}\right\rangle
\tea
Observe that this implies that $\left\langle \varphi^C\gamma^{DE}\right\rangle$ is totally symmetric.

\bea
\left\langle \Phi_H^C\Phi_H^D\Phi_H^E\Phi_H^F\right\rangle&=&\left\langle \left(\Phi_H^C\Phi_H^D\right)\left(\Phi_H^E\Phi_H^F\right)\right\rangle\nn
&=&\left(\phi^C\phi^D+G^{CD}\right)\left(\phi^E\phi^F+G^{EF}\right)\nn
&+&\phi^C\left[\phi^EG^{DF}+\phi^FG^{DE}+\left\langle \varphi^D\gamma^{EF}\right\rangle\right]\nn
&+&\phi^D\left[\phi^EG^{CF}+\phi^FG^{CE}+\left\langle \varphi^C\gamma^{EF}\right\rangle\right]\nn
&+&\phi^E\left\langle \varphi^F\gamma^{CD}\right\rangle
+\phi^F\left\langle \varphi^E\gamma^{CD}\right\rangle+\left\langle \gamma^{CD}\gamma^{EF}\right\rangle
\tea
We can now relate the derivatives of the mean fields with respect to the sources to stochastic averages

\be
\phi^{C,E}=i\left\langle \varphi^C\varphi^E\right\rangle
\te

\be
G^{CD,E}=i\left\langle \varphi^C\gamma^{DE}\right\rangle
\te

\be
\phi^{C,\left(EF\right)}=\frac i2\left\{\left\langle \varphi^C\gamma^{EF}\right\rangle+\phi^EG^{CF}+\phi^FG^{CE}\right\}
\te

\bea
G^{CD,\left(EF\right)}=\frac i2\left\{\phi^E\left\langle \varphi^F\gamma^{CD}\right\rangle
+\phi^F\left\langle \varphi^E\gamma^{CD}\right\rangle+\left\langle \gamma^{CD}\gamma^{EF}\right\rangle\right\}
\tea
Whereby we get the identities

\be
\left[\Gamma_{,AC}-2\Gamma_{,\left(AC\right)}\right]\left\langle \varphi^C\varphi^E\right\rangle+\Gamma_{,A\left(CD\right)}\left\langle \gamma^{CD}\varphi^E\right\rangle=i\delta_A^E
\te

\bea
&&\left[\Gamma_{,AC}-2\Gamma_{,\left(AC\right)}\right]\left\{\left\langle \varphi^C\gamma^{EF}\right\rangle+\phi^EG^{CF}+\phi^FG^{CE}\right\}\nn&+&\Gamma_{,A\left(CD\right)}\left\{\phi^E\left\langle \varphi^F\gamma^{CD}\right\rangle
+\phi^F\left\langle \varphi^E\gamma^{CD}\right\rangle+\left\langle \gamma^{CD}\gamma^{EF}\right\rangle\right\}=2i\delta_{\left(AB\right)}^{\left(EF\right)}\phi^B
\tea
which reduces to

\be
\left[\Gamma_{,AC}-2\Gamma_{,\left(AC\right)}\right]\left\langle \varphi^C\gamma^{EF}\right\rangle+\Gamma_{,A\left(CD\right)}\left\langle \gamma^{CD}\gamma^{EF}\right\rangle=0
\te

\be
\Gamma_{,\left(AB\right)C}\left\langle \varphi^C\varphi^E\right\rangle+\Gamma_{,\left(AB\right)\left(CD\right)}\left\langle \gamma^{CD}\varphi^E\right\rangle=0
\te

\be
\Gamma_{,\left(AB\right)C}\left\langle \varphi^C\gamma^{EF}\right\rangle+\Gamma_{,\left(AB\right)\left(CD\right)}\left\langle \gamma^{CD}\gamma^{EF}\right\rangle=\frac i2\left\{\delta_A^E\delta_B^F+\delta_A^F\delta_B^E\right\}
\te
Assuming the mean field equation $\Gamma_{,\left(AC\right)}=0$ these equation suggest a stochastic dynamics for the $\varphi$, $\gamma$ fields

\be
\Gamma_{,AC} \varphi^C+\Gamma_{,A\left(CD\right)}\gamma^{CD}=-\xi_A
\label{2pilan1}
\te
Observe that  a possible term $\kappa_{AB}\phi^B$ is absent,

\be
\Gamma_{,\left(AB\right)C} \varphi^C+\Gamma_{,\left(AB\right)\left(CD\right)} \gamma^{CD}=\frac {-1}2\kappa_{AB}
\label{2pilan2}
\te
provided

\be
\left\langle \xi_A\varphi^E\right\rangle=-i\delta_A^E
\te

\be
\left\langle \xi_A\gamma^{CD}\right\rangle=0
\te

\be
\left\langle \kappa_{AB}\varphi^E\right\rangle=0
\te

\be
\left\langle \kappa_{AB}\gamma^{EF}\right\rangle=-i\left\{\delta_A^E\delta_B^F+\delta_A^F\delta_B^E\right\}
\te
Multiplying the Langevin equations by the sources and using these expectation values, we get

\be
\left\langle \xi_A\xi_B\right\rangle=i\Gamma_{,AB}
\te

\be
\left\langle \xi_A\kappa_{CD}\right\rangle=2i\Gamma_{,A\left(CD\right)}
\te

\be
\left\langle \kappa_{CD}\kappa_{AB}\right\rangle=4i\Gamma_{,\left(AB\right)\left(CD\right)} 
\te

\subsection{Recovery of the 1PI stochastic theory from the 2PI one}

Let us check that the 2PI and 1PI theories agree as far as the mean field fluctuations are concerned. This must be true, as the 2PI theory is built around the requirement that $\left\langle \varphi^A\varphi^B\right\rangle=G^{AB}$

The 1PIEA $\Gamma_{1PI}$ is recovered from the 2PIEA $\Gamma$ as

\be
\Gamma_{1PI}\left[\phi\right]=\Gamma\left[\phi,G_0\left[\phi\right]\right]
\te
where the correlations $G_0$ are slaved to the mean field through

\be
\Gamma_{,\left(AB\right)}\left[\phi,G_0\left[\phi\right]\right]=0
\te
One further derivative shows that

\be
G^{CD}_{0,E}=-\left[\Gamma_{,\left(AB\right)\left(CD\right)}\right]^{-1}\Gamma_{,\left(AB\right)E}
\te
Therefore the first and second derivatives of the 1PIEA are

\be
\Gamma_{1PI,A}\left[\phi\right]=\Gamma_{,A}
\te

\be
\Gamma_{1PI,AB}\left[\phi\right]=\Gamma_{,AB}+\Gamma_{,A\left(CD\right)}G^{CD}_{0,B}
\te
The stochastic equation for the mean field fluctuations, as derived from the 1PIEA, is simply eq. (\ref{1pilan}) with noise self-correlation (\ref{1pinoise}). Now eq. (\ref{2pilan2}) admits the solution

\be
\gamma^{CD}=G^{CD}_{0,E}\varphi^E-\frac12\left[\Gamma_{,\left(AB\right)\left(CD\right)}\right]^{-1}\kappa_{AB}
\te
Using this into eq. (\ref{2pilan1}) we get back eq. (\ref{1pilan}) provided we identify

\be
\xi^{1PI}_A=\xi_A+\frac12G^{CD}_{0,A}\kappa_{CD}
\te
Indeed, both expressions have the same self-correlation.

\subsection{Stochastic equations for the physical propagators}

Finally, let us recover the noise self-correlation given in \cite{CalHu99}. The situation is similar to the one in the 1PI theory. In the 1PI theory, the $\phi^{-}$ field vanishes identically on-shell. However, in the stochastic approach we assign a nontrivial source $\xi^{1PI}_{-}$ to it. It is by eliminating this auxiliary field that we recover the usual approach, with a single stochastic source $\tilde{\xi}$ whose self-correlation is given by the noise kernel.

Similarly, in the quantum field theory problem the correlator $G^{--}=\left\langle \varphi^{-}\varphi^{-}\right\rangle$ vanishes identically, as a result of path ordering. However, in the stochastic approach, we consider it as an auxiliary field and couple a source to it. The authors of \cite{CalHu99} failed to recognize the violation of the constraint $G^{--}=0$, but compensated this oversight by forcing a sign change in their expression for the noise self-correlation. In the context of the proper theory, this sign change is due  to the elimination of the auxiliary field $G^{--}$, keeping only the physical degrees of freedom.

For simplicity, and as in \cite{CalHu99}, let us assume a symmetric scalar field theory, meaning that the background fields vanish and also $\Gamma_{,A\left(BC\right)}=0$. Then the 2PI EA can be written as

\be
\Gamma=\frac12S_{AB}\mathbf{G}^{AB}-\frac i2\mathrm{Tr}\:\mathrm{Ln}\left[\mathbf{G}\right]+\Gamma_Q
\te

The first derivatives of the 2PI CTP EA yield the mean field equations of motion

\be
S_{AB}-i\mathbf{G}^{-1}_{AB}+2\Gamma_{Q,\left(AB\right)}=0
\te
we shall call ${G}^{AB}$ the on-shell propagators. We adopt the $\left(\pm ,a\right)$ indexes, as before. In this representation, we have the identifications eq. (\ref{onshellprop}) and (\ref{onshelleq}). We now replace the generic kernels $\mathbf{G}^{AB}$ by the stochastic kernels ${G_s}^{AB}={G}^{AB}+\gamma^{AB}$ and expand the effective action to second order. For simplicity, we only keep linear terms in $\Gamma_Q$. We show in \cite{CalHu99} that this is enough to study the fluctuation terms in the kinetic field theory limit. The relevant quadratic terms in the effective action are

\be
\Gamma^{\left(2\right)}=\frac i4\mathrm{Tr}\left[{G}^{-1}\gamma\right]^2
\te
where, as in eq. (\ref{onshelleq})

\be
{G}^{-1}=\left(\begin{array}{cc}
	0&-i\mathbf{D}_{b,a}\\
-i\mathbf{D}_{a,b}&\mathbf{N}_{ab}
\end{array}\right)
\te
therefore

\be
{G}^{-1}\gamma =\left(\begin{array}{cc}
	-i\mathbf{D}_{b,a}\gamma^{\left(-b\right)\left(+c\right)}&-i\mathbf{D}_{b,a}\gamma^{\left(-b\right)\left(-c\right)}\\
-i\mathbf{D}_{a,b}\gamma^{\left(+b\right)\left(+c\right)}+\mathbf{N}_{ab}\gamma^{\left(-b\right)\left(+c\right)}&-i\mathbf{D}_{a,b}\gamma^{\left(+b\right)\left(-c\right)}+\mathbf{N}_{ab}\gamma^{\left(-b\right)\left(-c\right)}
\end{array}\right)
\te
and

\bea
\Gamma^{\left(2\right)}&=&\frac {-i}4\left\{\mathbf{D}_{b,a}\gamma^{\left(-b\right)\left(+c\right)}\mathbf{D}_{d,c}\gamma^{\left(-d\right)\left(+a\right)}\right.\nn
&+&\left.2\mathbf{D}_{b,a}\gamma^{\left(-b\right)\left(-c\right)}\left[\mathbf{D}_{c,d}\gamma^{\left(+d\right)\left(+a\right)}+i\mathbf{N}_{cd}\gamma^{\left(-d\right)\left(+a\right)}\right]\right.\nn
&+&\left.\left[\mathbf{D}_{a,b}\gamma^{\left(+b\right)\left(-c\right)}+i\mathbf{N}_{ab}\gamma^{\left(-b\right)\left(-c\right)}
\right]\left[\mathbf{D}_{c,d}\gamma^{\left(+d\right)\left(-a\right)}+i\mathbf{N}_{cd}\gamma^{\left(-d\right)\left(-a\right)}
\right]\right\}
\tea
The stochastic equations for the propagators are

\be
\frac {-i}2\mathbf{D}_{b,a}\left[\gamma^{\left(-b\right)\left(+c\right)}\mathbf{D}_{d,c}+i\gamma^{\left(-b\right)\left(-c\right)}\mathbf{N}_{cd}\right]=\frac{-1}2\kappa_{\left(-d\right)\left(+a\right)}
\te

\be
\frac {-i}2\left[\mathbf{D}_{a,b}\gamma^{\left(+b\right)\left(-c\right)}+i\mathbf{N}_{ab}\gamma^{\left(-b\right)\left(-c\right)}
\right]\mathbf{D}_{c,d}=\frac{-1}2\kappa_{\left(+d\right)\left(-a\right)}
\te

\bea &&\frac{-i}2\left\{\mathbf{D}_{b,a}\left[\mathbf{D}_{c,d}\gamma^{\left(+d\right)\left(+a\right)}+i\mathbf{N}_{cd}\gamma^{\left(-d\right)\left(+a\right)}\right]\right.\nn
&+&\left.i\mathbf{N}_{ab}\left[\mathbf{D}_{c,d}\gamma^{\left(+d\right)\left(-a\right)}+i\mathbf{N}_{cd}\gamma^{\left(-d\right)\left(-a\right)}
\right]\right\}=\frac{-1}2\kappa_{\left(-b\right)\left(-c\right)}
\tea

\be
\frac{-i}2\mathbf{D}_{b,a}\gamma^{\left(-b\right)\left(-c\right)}\mathbf{D}_{c,d}=\frac{-1}2\kappa_{\left(+d\right)\left(+a\right)}
\te
This last equation implies

\be
\gamma^{\left(-b\right)\left(-c\right)}=-iG_{adv}^{cd}G_{adv}^{ba}\kappa_{\left(+d\right)\left(+a\right)}
\te
Observe that $\gamma^{\left(-b\right)\left(-c\right)}$ is not zero in the equivalent stochastic problem. 
The equations for the physical stochastic propagators are obtained eliminating $\gamma^{\left(-b\right)\left(-c\right)}$ throughout. For example, for the fluctuations in the Hadamard propagator we obtain

\be
\mathbf{D}_{c,d}\gamma^{\left(+d\right)\left(+a\right)}\mathbf{D}_{b,a}+i\mathbf{N}_{cd}\gamma^{\left(-d\right)\left(+a\right)}\mathbf{D}_{b,a}+i\mathbf{D}_{c,d}\gamma^{\left(+d\right)\left(-a\right)}\mathbf{N}_{ab}=-i\tilde{\kappa}_{\left(-b\right)\left(-c\right)}
\te
where

\bea
\tilde{\kappa}_{\left(-b\right)\left(-c\right)}&=&\kappa_{\left(-b\right)\left(-c\right)}+i\mathbf{N}_{cd}\gamma^{\left(-d\right)\left(-a\right)}\mathbf{N}_{ab}\nn
&=&{\kappa}_{\left(-b\right)\left(-c\right)}+\mathbf{N}_{ce}G_{adv}^{fd}G_{adv}^{ea}\mathbf{N}_{fb}\kappa_{\left(+d\right)\left(+a\right)}
\tea
From the self-correlations of the original sources

\be
\left\langle \kappa_{\left(-b\right)\left(-c\right)}\kappa_{\left(-d\right)\left(-a\right)}\right\rangle=-2\mathbf{N}_{ab}\mathbf{N}_{cd}
\label{ne101}
\te

\be
\left\langle \kappa_{\left(-b\right)\left(-c\right)}\kappa_{\left(+d\right)\left(+a\right)}\right\rangle=2\mathbf{D}_{b,a}\mathbf{D}_{c,d}
\te

\be
\left\langle \kappa_{\left(+b\right)\left(+c\right)}\kappa_{\left(+d\right)\left(+a\right)}\right\rangle=0
\te
we get

\be
\left\langle \tilde{\kappa}_{\left(-b\right)\left(-c\right)}\tilde{\kappa}_{\left(-d\right)\left(-a\right)}\right\rangle=2\mathbf{N}_{ab}\mathbf{N}_{cd}
\label{ne102}
\te
which agrees with \cite{CalHu99}. Observe that in going from eq. (\ref{ne101}) to eq. (\ref{ne102}) the sign of the right hand side has changed. Thus we obtain the sign change enforced in \cite{CalHu99}, whose origin is now clear.

\section{Final remarks}

In this paper we have presented a systematic stochastic approach which allows to build Langevin-like generalizations of the Schwinger-Dyson equations of an interacting quantum field theory. The resulting fluctuations account both for quantum and statistical fluctuations.

It is hoped that the present approach will prove clearer and easier to apply than existing alternatives in the literature. Quantum and statistical fluctuations are increasingly under scrutiny in fields that range from cosmology to Bose-Einstein condensates. We expect the tool we have presented in this note will prove valuable in this endeavor. 

\ack

This work is supported in part by ANPCyT, CONICET and UBA, Argentina.

\end{document}